\documentclass{iopart}
\usepackage{psfig}

\newcommand{\forget}[1]{\iffalse#1\fi}
\newcommand{\forgetmenot}[1]{\iftrue#1\fi}

\newcommand{\be}{\begin{equation}}
\newcommand{\ee}{\end{equation}}
\newcommand{\ba}{\begin{eqnarray}}
\newcommand{\ea}{\end{eqnarray}}

\renewcommand{\;}[2]{{\frac{#1}{#2}}}
\renewcommand{\;}[2]{{\frac{#1}{#2}}}
\newcommand{\bra}[1]{\left(#1\right)}
\newcommand{\bras}[1]{\left[#1\right]}
\newcommand{\brac}[1]{\left\{#1\right\}}

\newcommand{\lapp}{\mathrel{\vcenter{\hbox{\tiny \ooalign{\raise 3.25pt
        \hbox{$<$}\crcr $\sim$}}}}}
\newcommand{\gapp}{\mathrel{\vcenter{\hbox{\tiny \ooalign{\raise 3.25pt
        \hbox{$>$}\crcr $\sim$}}}}}
\newcommand{\eqdef}{\!\!\mathrel{\vcenter{\hbox{ \ooalign{\raise 4.75pt
        \hbox{${\textsf{\tiny{def}}}$}\crcr $=$}}}}}

\newcommand{\del}{\nabla}
\newcommand{\sdel}{{\mathrm{D}}}

\renewcommand{\>}{\rangle}
\renewcommand{\div}{{\mathsf{div}}\,}
\newcommand{\curl}{{\mathsf{curl}}\,}

\newcommand{\mur}{\mu}
\newcommand{\mum}{\rho_{\mathrm{B}}}
\newcommand{\mud}{\rho_{\mathrm{M}}}
\def\e{\epsilon}
\newcommand{\ed}{\epsilon^\dag}
\newcommand{\eds}{\epsilon^{\dag\star}}
\newcommand{\edss}{\epsilon^{\dag\star\star}}
\newcommand{\edsss}{\epsilon^{\dag\star\star\star}}
\newcommand{\edd}{\epsilon^{\dag\dag}}
\newcommand{\edds}{\epsilon^{\dag\dag\star}}
\newcommand{\eddd}{\epsilon^{\dag\dag\dag}}
\newcommand{\edddd}{\epsilon^{\dag\dag\dag\dag}}
\newcommand{\eddds}{\epsilon^{\dag\dag\dag\star}}
\newcommand{\eddss}{\epsilon^{\dag\dag\star\star}}
\def\es{\epsilon^{\star}}
\newcommand{\ess}{\epsilon^{\star\star}}
\newcommand{\esss}{\epsilon^{\star\star\star}}

\begin{document}

\title{CMB limits on large-scale magnetic fields\\
in an inhomogeneous universe}

\author{C. A. Clarkson$^{1,2}$\footnote{clarkson@maths.uct.ac.za},
A. A. Coley$^1$\footnote{aac@mathstat.dal.ca}, R.
Maartens$^3$\footnote{roy.maartens@port.ac.uk}, C. G.
Tsagas$^{3,2}$\footnote{ctsagas@maths.uct.ac.za}}
\address{~}
\address{$^1$ Department of Mathematics \& Statistics,
Dalhousie University, Halifax~B3H~3J5, Canada}
\address{~}
\address{$^2$ Relativity \& Cosmology Group, Department of
Mathematics \& Applied Mathematics, University of Cape Town, Cape
Town~7701, South Africa}
\address{~}
\address{$^3$ Institute of Cosmology \& Gravitation, University of
Portsmouth, Portsmouth~PO1~2EG, UK}



\begin{abstract}

We use the cosmic microwave background temperature anisotropy to
place limits on large-scale magnetic fields in an inhomogeneous
(perturbed Friedmann) universe. If no assumptions are made about
the spacetime geometry, only a weak limit can be deduced directly
from the CMB. In the special case where spatial inhomogeneity is
neglected to first order, the upper limit is much stronger, i.e. a
few~$\times 10^{-9}\,$G.

\end{abstract}

\section{Introduction}

Magnetic fields have been observed in the universe on a wide range
of scales. Fields with strengths of a few $\mu$G are prolific in
galaxies and galaxy clusters, extending well beyond the core
regions of the latter, and have also been detected in high
redshift Lyman-$\alpha$ objects. Magnetic fields in extragalactic
structures are detected mainly via radio polarisation studies,
X-ray emission and Faraday rotation measurements (see~\cite{K} for
a comprehensive review). Magnetic fields in galaxies and galaxy
clusters appear to be the result of the nonlinear amplification of
weak seed fields, mainly via the galactic dynamo. However, the
detection of ordered magnetic fields in high redshift objects
(with $z>2$) poses a stiff challenge to the dynamo mechanism. As
yet, there is no direct evidence of magnetic field presence on
cosmological scales, corresponding to a significant fraction of
the Hubble length. Clearly, any such field could not arise through
structure formation physics, but it would have to be the remnant
of a primordial field, redshifting with expansion:
\begin{equation}
B=B_0\left({a_0\over a}\right)^2,
\end{equation}
where $B_0$ is the current field strength.

The strength of primordial, cosmological magnetic fields is
limited by observed helium abundances and by the near-isotropy of
the cosmic microwave background (CMB) (see~\cite{GR1} for recent
reviews). Any magnetic field present at the time of cosmological
nucleosynthesis inevitably affects the abundance of primordial
Helium, since it provides an additional form of relativistic
energy density. This, in turn, increases the expansion rate of the
universe with the effect that the neutron-proton freeze out of
weak interactions occurs at a higher temperature. The result is an
increase in the synthesised abundance of primordial Helium. Hence,
Helium-4 observations (extrapolated to zero metalicity) provide an
upper limit of $\sim10^{-7}G$, in today's values, on any
primordial magnetic field present at nucleosynthesis~\cite{GR2}.

Stronger limits are imposed from the observed high isotropy of the
CMB photons. The COBE data place an upper bound on a homogeneous
magnetic field present at the time of last scattering. In a recent
analysis of a particular class of spatially homogeneous Bianchi
universes, an upper bound of $B_0\lapp 10^{-9}\,$G was
obtained~\cite{BFS}. Here we generalize previous work to the case
of inhomogeneous fields in an inhomogeneous almost-Friedmann
universe. We also generalize the limits found in~\cite{BFS} by
weakening some of their assumptions. It turns out that, if we do
not assume a spatially homogeneous geometry, the limits imposed
directly by CMB data on super-Hubble magnetic fields in an
inhomogeneous universe are much weaker, $B_0\lapp 10^{-6}\,$G. A
similar situation arises when considering the limits placed on the
shear by CMB anisotropies, as pointed out in~\cite{MES3}.

Recently it was proven~\cite{CC} under quite general circumstances
that a magnetic field is prohibited in spacetimes where exactly
isotropic radiation is also present. Taking this as our starting
point, small anisotropy allows for a weak magnetic field. We use
the 1+3-covariant analysis of CMB temperature
anisotropies~\cite{SME,MES,CL} and of magnetic fields~\cite{TBM},
in order to derive limits on large-scale fields as a function of
coherence scale. Following the approach of~\cite{MES,MES2}, we use
the radiation multipoles to derive limits which are
model-independent, in the sense that they do not rely on
assumptions about the (perturbed Friedmann) spacetime geometry.

In Sec.~\ref{EGSsec} we outline the general formalism for imposing
limits on large-scale magnetic fields from observed CMB
temperature anisotropies. In Sec.~\ref{thelimits}, we give our
main results, which follow from a refinement of the method
in~\cite{MES2}. For convenience, we omit most of the calculational
details, and give the key equations in Appendices~A and B. We use
units such that $8\pi G=1=c$. Our notation follows that
of~\cite{MGE}. In particular, $\dot X_{a\cdots b}= u^c\del_c
X_{a\cdots b}$ and $\sdel_a$ is the covariant derivative in the
rest space, i.e., $\sdel_c X_{a\cdots b}= h_{c}^{~d}
h_a^{~e}\cdots h_b^{~f}\del_d X_{e\cdots f}$, where $h_{ab}=
g_{ab}+u_au_b$ is the projection tensor. Angled brackets on
indices denote the projected, symmetric and trace free (PSTF)
part. The 3-divergence and 3-curl of PSTF tensors are $\div
X_{a}=\sdel^b X_{ab}$, $\cdots$, and $\curl X_a=\varepsilon_{abc}
\sdel^bX^c$, $\curl X_{ab}=\varepsilon_{cd(a}\sdel^c
X_{b)}{}{}^d$, $\cdots$.

\section{CMB anisotropy induced by large-scale
magnetic fields}\label{EGSsec}

In the 1+3-covariant analysis of CMB anisotropy~\cite{MES,CL}, a
physical choice of 4-velocity $u^a$ is made, usually the
4-velocity of cold dark matter (CDM), and all perturbative
quantities are then covariant vectors or tensors in the rest-space
of $u^a$, with direct geometrical or physical meaning. The
fractional temperature fluctuation is expanded in covariant
multipoles $\tau_{A_\ell}$ ($A_\ell = a_1\cdots a_\ell$), which
are PSTF tensors. These are limited directly by observations:
\begin{equation}
|\tau_{A_\ell}|\equiv \sqrt{\tau_{A_\ell}\tau^{A_\ell}}<\e_\ell.
\end{equation}
COBE data leads to the values~\cite{SAG}
 \ba
\e_2&\approx& 1.1(\pm 0.8)\times 10^{-5}\,,\label{e1}\\
\e_3&\approx& 2.5(\pm 1.3)\times 10^{-5}\,,\label{e2}
 \ea
which we use here. The first moment $\tau_a$ is the dipole, which
is usually attributed to our peculiar motion relative to the CMB
frame. We assume this motion is corrected for by setting
$\tau_a=0=\e_1$ for the bulk of the paper. However, it is possible
that a residual dipole of cosmological origin exists (it would be
frequency dependent, and thus could not be set to zero by a
Lorentz boost), and we include it in our calculations for
generality.

In addition to the observed bounds on the $\tau_{A_\ell}$, we need
bounds on their temporal and spatial gradients, in order to find
limits on geometrical and physical quantities that characterize
the spacetime. We define the expansion-normalized, dimensionless
$\epsilon$-quantities~\cite{MES,MES2}
 \ba
|\dot\tau_{A_\ell}|&<& 3H \es_\ell,~~~|\ddot\tau_{A_\ell}|< 9H^2
\ess_\ell,~~~\cdots,\nonumber\\ |\sdel_a\tau_{A_\ell}|&<& 3H
\ed_\ell,~~~ |\sdel_a\sdel_b\tau_{A_\ell}| < 9H^2
\edd_\ell,~~~\cdots,\nonumber\\
|(\sdel_a\tau_{A_\ell})^{\displaystyle\cdot}|&<& 9H^2
\eds_\ell,~~~ |(\sdel_a\sdel_b\tau_{A_\ell})^{\displaystyle\cdot}|
< 27H^3 \edds_\ell,~~~\cdots,\label{boundsIV}
 \ea
where $H$ is the background Hubble rate.

Following~\cite{MES2}, we can find upper bounds on all
perturbative quantities in terms of the
$\e_\ell,~\es_\ell,~\ed_\ell$, etc. However, the derivative bounds
$\es_\ell,~\ed_\ell\cdots$ are not directly measurable, as we are
unable to move a cosmological time- or space-separation from our
current spacetime event. Here we make the simple assumption that
the time- and space-variations of the multipoles are governed
respectively by the Hubble rate and the physical scale $\lambda$
of the perturbation, i.e.,
 \be
\left|\dot\tau_{A_\ell}\right|\sim H|{\tau_{A_\ell}}|,~~ \left|
\sdel_a\tau_{A_\ell}\right|\sim\;1
\lambda|{\tau_{A_\ell}}|.\label{grads}
 \ee
This assumption implies
 \ba
\es_\ell\sim {1\over3}\e_\ell, ~~~\ess_\ell\sim {1\over9}
\e_\ell,~~~\cdots,\label{estar}\\ \ed_\ell\sim {1\over 3\beta}
\e_\ell,~~~\edd_\ell\sim\;{1} {9\beta^2}\e_\ell,~~~\cdots
~~~~\mbox{where}~ \beta={\lambda\over H^{-1}}. \label{edag}
 \ea
The dimensionless parameter $\beta$ gives the coherence scale as a
fraction of the Hubble length, with $\beta\gapp O(1)$ since we are
considering large scales.

The magnetic field is `frozen' into the baryonic fluid, which may be treated as
an infinitely conducting medium. Here, we neglect the peculiar velocity of CDM
relative to the baryons. The physical justification for doing so comes from the
fact that we address the linear regime and consider large scales, where the
velocity difference between the two components is expected to be minimal. On
these grounds, we choose $u^a$ to be the 4-velocity of CDM-baryon fluid with
total density $\mud$. The kinematics of the pressure-free matter are
characterized by the volume expansion $\Theta$, rotation $\omega^a$,
acceleration $A^a$ and shear distortion $\sigma_{ab}$ of $u^a$. Note that, even
in the absence of pressure gradients, the flow lines are generally non-geodesic
(i.e.~$A_a\neq0$) due to the magnetic field presence. Here, however, we will
assume an effectively force-free field (i.e.~$\varepsilon_{abc}B^b{\rm
curl}B^c=0$), and ignore the acceleration to first order. This is a reasonable
approximation, given that the field is too weak to affect the motion of the
baryonic matter. It will also allow us to focus upon the purely anisotropic
magnetic effects.

The energy-momentum tensor of the magnetized dust is
\begin{equation}
T_{ab}=(\mud+\mum)u_au_b+ p_{\rm B}h_{ab}+ \Pi_{ab}\,, \label{Tab}
\end{equation}
where $\mum=B^2/8\pi$ and $p_{\rm B}=\mum/3$ are the magnetic
energy density and isotropic pressure respectively. Also,
$\Pi_{ab}=-B_{\langle a}B_{b\rangle}/4\pi$ is the symmetric and
trace free tensor that conveys the anisotropic magnetic effects.
The radiation energy-momentum tensor is
\begin{equation}
{\cal T}_{ab}=\mur u_au_b+ \;13\mur h_{ab}+ 2u_{(a}q_{b)}+
\pi_{ab}\,, \label{cTab}
\end{equation}
where $\mur$,  $q_a$ and $\pi_{ab}$ are the photon energy density,
momentum density and anisotropic stress. These are directly
related to the temperature anisotropy multipoles by~\cite{MES}
 \be
q_a=\;43{\mur\tau_a},~~~\pi_{ab}=\;{8}{15}
\mur\tau_{ab}.\label{q->tau}
 \ee
The photon energy momentum tensor involves only the first two
multipoles, but we will require also the octupole
 \be
\xi_{abc}=\;{8}{35}\mur\tau_{abc},\label{xi->tau}
 \ee
which appears in the evolution equation for $\pi_{ab}$,
Eq.~(\ref{dotpiab}).

The field equations $G_{ab}=T_{ab}+{\cal T}_{ab}$, the Ricci
identities and the Bianchi identities may be split into a set of
evolution (along $u^a$) and constraint equations. The evolution of
the magnetic field is determined by Maxwell's equations. The
reader is referred to Appendix~A for the necessary equations.

\section{The limits}\label{thelimits}

We first present the CMB limits on inhomogeneous magnetic fields
as a function of coherence scale, $\lambda$; then we discuss the
homogeneous case.

\subsection{Inhomogeneous universe}\label{inhomo}

Our procedure to find constraints on the magnetic field strength
$\mum$, or equivalently $|\Pi_{ab}|$, is a generalization of the
non-magnetized analysis in~\cite{MES2}. Briefly, we manipulate the
field equations to express $\Pi_{ab}$ in terms only of the
radiation quantities $\mur, q_a, \pi_{ab}, \xi_{abc}$. This is
facilitated by the appearance of the shear in Eq.~(\ref{dotpiab}),
which, in the absence of acceleration, is the only coupling of the
radiation to the first-order kinematical quantities. The main
aspects of this calculation are in Appendix~B, and the key result
is Eq.~(\ref{pilims}).

Neglecting the dipole moment and the energy density of the
radiation~$\Omega_\mathrm{R}$, and restoring units,
Eq.~(\ref{pilims}) gives
 \ba
B_0&<&\max(B)\equiv\bra{\;32}^{3/4}\;{cH_0}{\sqrt
G}\left\{\;14\Omega_\mathrm{M}\bra{5\es_2+\;{45}{7}\ed_3}
    +\Omega_\Lambda\bra{\es_2+\;97\ed_3}\right.\nonumber\\
    &+&\left.2\es_2+\;{15}{2}\ess_2
    +\;92\esss_2+3\edd_2+\;92\edds_2+
    \;{81}{10}\edddd_2\right.\nonumber\\
    &+&\left. \;{18}{7}\ed_3+\;{135}{14}\eds_3+\;{81}
    {14}\edss_3
    +\;{81}{14}\eddd_3 \right\}^{1/2},\label{Blims}
 \ea
where the $\epsilon$'s are evaluated at the current time. The
function $\max(B)$ gives upper limits on large-scale magnetic
fields, coherent on a given scale $\lambda$, imposed by CMB
temperature anisotropies. This upper limit is given directly in
terms of CMB multipoles and their derivatives, and is
model-independent, i.e., no assumptions have been made about the
spacetime geometry.

\begin{figure}[h!]
\protect\centerline{\psfig{file=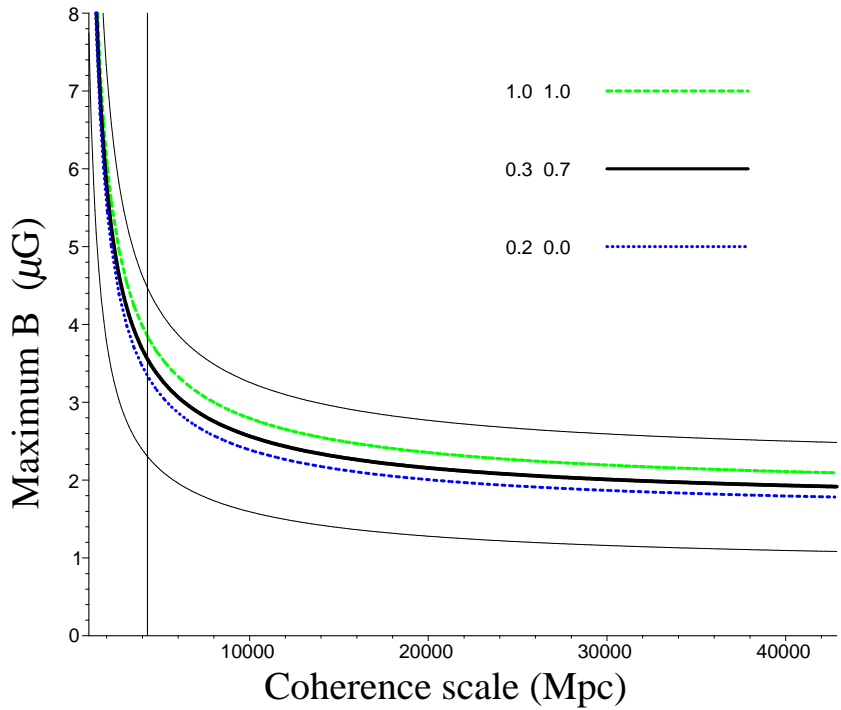,width=11cm,angle=0}}
\caption{Maximum magnetic field strength in $\mu$G on large scales
as a function of coherence scale~$\lambda$ ($\gapp 1000$~Mpc) for various
cosmological parameters~$\{\Omega_\mathrm{M},\Omega_\Lambda\}$ shown in the
key, with $h=H_0/(100~\mbox{km/s/Mpc})=0.7$. The concordance
model~$\{\Omega_\mathrm{M}, \Omega_\Lambda\}=\{0.3,0.7\}$ (central thick solid
curve) is straddled on either side by the collective uncertainty in $\e_2$ and
$\e_3$ (outer thin solid curves). An open and a closed model with extreme
parameters are shown by the dashed curves, indicating how little the
cosmological parameters affect the upper limit, and certainly much less than
the observational uncertainty. The vertical line represents the Hubble
distance, $\lambda=3000h^{-1}$~Mpc. Because our model is only valid down to
scales of the order of the Hubble distance, scales significantly below this
should be interpreted with care.}
\label{maxBfig}
\end{figure}

For a numerical estimate, we need to use the simple assumptions of
Eqs.~(\ref{estar}) and (\ref{edag}), in order to evaluate the
derivative-$\epsilon$'s. Then we find, in terms of the observable
quantities $\epsilon_2, \epsilon_3$, that
 \ba\label{Blims2}
\max(B)&=&\bra{\;32}^{3/4}\;{cH_0}{\sqrt G}\left\{
\e_2\bras{\;{5}{12}\Omega_\mathrm{M}+\;13\Omega_{\Lambda}+\;53
+\;{1}{2}\beta^{-2}+ \;{1}{10}\beta^{-4}} \right.\nonumber\\
&&~\left.
+\e_3\bras{\;{1}{7}\bra{\;54\Omega_\mathrm{M}+\Omega_\Lambda +5
}\beta^{-1} +\;{3}{14}\beta^{-3}} \right\}^{1/2}.
 \ea
This is our main result. Using the limits in Eqs.~(\ref{e1}) and
(\ref{e2}), we can evaluate max($B$), which is plotted in
Fig.~\ref{maxBfig}. One of the main features of the plot is that
uncertainty in $\e_2$ and $\e_3$ from COBE data produces a far
greater uncertainty than the uncertainty in the cosmological
parameters.

On the largest scales, $\beta\to\infty$, Eq.~(\ref{Blims})
simplifies to give
\begin{equation}\label{hom}
\max(B)\big|_{\beta \to\infty} =\bra{\;32}^{3/4}\;{cH_0}{\sqrt G}
\left\{ {\;{5}{12}\Omega_\mathrm{M}+\;13\Omega_{\Lambda}+\;53 }
 \right\}^{1/2}\e_2\sim 10^{-6}{\rm G}.
\end{equation}

\subsection{Spatially homogeneous universe}\label{homo}

The upper limit on $B_0$ on the largest scales in the general case
of an inhomogeneous universe, as given by Eq.~(\ref{hom}), is much
weaker than the limit that can be imposed if we assume that the
universe is spatially homogeneous to first order. Homogeneity
implies that we can set to zero the $\epsilon$'s that involve
gradients, as we did in deriving Eq.~(\ref{hom}).\footnote{%
Note that homogeneous radiation multipoles, i.e.
$\sdel_a\tau_{A_\ell}=0$, imply a Bianchi spacetime to first
order~\cite{MES}.%
} However, it is not only the radiation multipoles that are
homogeneous to first order, but the whole spacetime, leading to a
Bianchi model. The special dynamics of Bianchi models then leads
to a tighter constraint on $B_0$. A similar situation arises when
deriving limits on the shear $\sigma_{ab}$~\cite{MES3}.

It follows from~\cite{MES2} that the spatial 3-curvature vanishes
to first order, ${\cal R}_{ab}={\cal R}=0$. In addition the shear
becomes, from Eq.~(\ref{dotpiab}),
 \be
 \sigma_{ab}=-\dot\tau_{ab}.
  \ee
Thus the shear evolution equation becomes, using Eq.~(\ref{GC}),
 \be
\Pi_{ab}=\ddot\tau_{ab}+\Theta\dot\tau_{ab}
+\;{8}{15}\mur\tau_{ab}\label{Pi=pi}.
 \ee
The magnetic field acts as a forcing term for the quadrupole. The
particular solution associated with this forcing term is
 \be\label{pi}
\Pi_{ab}=\;{8}{15}\mur\tau_{ab},
 \ee
with $\dot\tau_{ab}=0$ for the particular solution. The solution
to the (non-magnetic) homogeneous part oscillates (at frequency
$\approx \sqrt{{8}\mur/15}$), while being suppressed with a
damping scale of the Hubble time. Thus a conservative upper limit
is given by Eq.~(\ref{pi}); using Eq.~(\ref{e2}) and
$\Omega_\mathrm{R}\sim2.5 h^{-2}\times10^{-5}\Omega_{\mathrm{M}}$,
we find that
 \be
B_0<6.2^{+1.9}_{-3.0}\times10^{-9} \sqrt{\Omega_{\mathrm{M}}}~G.
\label{BFSresult}
 \ee
With $\Omega_{\mathrm{M}}=0.3$, Eq.~(\ref{BFSresult}) gives
 \be
B_0<3.4^{+1.0}_{-1.6}\times10^{-9}G.
 \ee

This confirms the value found in~\cite{BFS}, and is derived under
slightly weaker assumptions; the spacetime is not chosen as a
specific exact Bianchi model, but is homogeneous to first order,
and turns out to be Bianchi~I if we start with a flat Friedmann
background. Furthermore, we include both the radiation energy
density and the cosmological constant. In the set of Bianchi
models which admit a pure magnetic field (types I, II, III,
IV$_o$, VII$_o$), they are all of the same genericity; therefore
we may consider this more general than~\cite{BFS}, where the
geometry is assumed to be type~VII$_h$.

\section{Conclusions}\label{conclusions}

For large-scale magnetic fields in an inhomogeneous
almost-Friedmann universe, we have found upper limits on the field
strength directly in terms of the CMB temperature multipoles and
their derivatives, as given by Eqs.~(\ref{Blims}) and
(\ref{Blims2}). On super-Hubble scales, this upper limit is very
weak:
\[
B_0 \lapp 2 \times 10^{-6}~{\rm G}
\]
for the concordance model, as shown by Eq.~(\ref{hom}) and Fig.~1.

When the almost-Friedmann universe is assumed to be homogeneous to
first order, i.e. a Bianchi spacetime, the upper limit is much
stricter, as given by Eq.~(\ref{BFSresult}). This generalizes the
result of~\cite{BFS}, by including a cosmological constant and
removing initial assumptions of a choice of model.

These limits have been derived in a covariant and gauge invariant
way using the 1+3 formalism. A major feature of our approach is
that our limits are largely model-independent, being derived from
properties of the Einstein-Boltzmann equations. Our main
assumptions are imposed on the $\epsilon$-quantities that bound
the derivatives of radiation multipoles, which are in principle
observable but in practice are not measurable.

It is also possible to find limits on the inhomogeneity of the
magnetic field, as given by the gradient of the magnetic energy
density $\sdel_a\mum$. The result is given in Appendix~B by
Eq.~(\ref{Dmub}).

\ack

CAC was supported by NSERC (at Dalhousie) and NRF (at Cape Town). AAC
acknowledges funding from NSERC. RM was supported by PPARC. CGT was supported
by PPARC (at Portsmouth) and Sida/NRF (at Cape Town). RM and CGT thank
Dalhousie University for hospitality while part of this work was done.

\appendix

\section{The linearized equations}\label{1+3eqns}

We assume that CDM and baryons share the same 4-velocity, which
coincides with that of the fundamental observers. Also, confining
ourselves to times after last scattering, we may treat the
magnetized dust and the radiation fluid as independently conserved
entities (i.e., ${\del}^bT_{ab}=0={\del}^b{\cal T}_{ab}$). Then we
arrive at the following linearized evolution equations for the
magnetised dust (see~\cite{TBM})
\begin{eqnarray}
\dot\mud+ \Theta\mud&=&0\,,  \label{dotrhod}\\ \mud A_a+
{{1\over3}}{\rm D}_a\mum+ \div\Pi_{a}&=&0\,, \label{Aa}\\
\dot\mum+ {{4\over3}}\Theta\mum&=&0\,, \label{dotB2}\\
\dot{\Pi}_{ab}+ {{4\over3}}\Theta\Pi_{ab}&=&0\,, \label{dotPiab}
\end{eqnarray}
and
\begin{eqnarray}
\dot\mur+ {{4\over3}}\Theta\mur+ \div q&=&0\,, \label{dotrhor}\\
\dot{q}_a+ {{4\over3}}\Theta q_a+ {{4\over3}}\mur A_a+
{{1\over3}}{\rm D}_a\mur+ \div\pi_{a}&=&0\,, \label{dotqa}\\
\dot{\pi}_{ab}+ {{4\over3}}\Theta\pi_{ab}+
{{8\over15}}\mur\sigma_{ab}+ 2{\rm D}_{\langle a}q_{b\rangle}+
\div\xi_{ab}&=&0\,, \label{dotpiab}
\end{eqnarray}
for the photons (see \cite{MGE}). Note that in deriving
Eq.~(\ref{Aa}) we have used the linear relation
$\varepsilon_{abc}B^b\curl B^c={{1\over6}}{\rm D}_aB^2+{\rm
D}^b\Pi_{ab}$ (recall that ${\rm D}^aB_a=0$). To first order, the
kinematic evolution is given by
\begin{eqnarray}
\dot{\Theta}+ {{1\over3}}\Theta^2+ {{1\over2}}(\mud+2\mur+2\mum)-
\div A-\Lambda&=&0\,, \label{Ray}\\ \dot{\sigma}_{ab}+
{{2\over3}}\Theta\sigma_{ab}+ E_{ab}- {{1\over2}}\pi_{ab}-
{{1\over2}}\Pi_{ab}- {\rm D}_{\langle a}A_{b\rangle}&=&0\,,
\label{dotsigma}\\ \dot{\omega}_a+ {{2\over3}}\Theta\omega_a+
{{1\over2}}{\curl}A_a&=&0\,,  \label{dotomega}\\ q_a-
{{2\over3}}{\rm D}_a\Theta+ \div\sigma_{a}- \curl \omega_a&=&0\,,
\label{con1}\\ \div\omega&=&0\,, \label{con2}
\end{eqnarray}
where $A_a$ is given by Eq.~(\ref{Aa}). Finally, the spacetime
geometry is determined by~\cite{MGE}
\begin{eqnarray}
&&\dot{E}_{ab}+ \Theta E_{ab}- \curl H_{ab}+
{{1\over2}}(\mud+{{4\over3}}\mur)\sigma_{ab}+\nonumber\\ &&~~~~~~
{{1\over2}}\dot{\pi}_{ab}+ {{1\over6}}\Theta\pi_{ab}-
{{1\over2}}\Theta\Pi_{ab}+ {{1\over2}}{\rm D}_{\langle
a}q_{b\rangle}=0\,,  \label{dotEab}\\&& \dot{H}_{ab}+ \Theta
H_{ab}+ {\curl}E_{ab}- {{1\over2}}\curl \pi_{ab}-
{{1\over2}}{\curl}\Pi_{ab}=0\,, \label{dotHab}\\&& H_{ab}-
\curl\sigma_{ab}- {\rm D}_{\langle a}\omega_{b\rangle}=0\,,
\label{con3}\\ &&\div E_{a}+ {{1\over2}}\div\pi_{a}+
{{1\over2}}\div\Pi_{a}- {{1\over3}}{\rm D}_a(\mud+\mur+\mum)+
{{1\over3}}\Theta q_a =0\,, \label{con4}\\&& \div H_{a}+
{{1\over2}}\curl q_a- (\mud+{{4\over3}}\mur)\omega_a=0\,,
\label{con5}\\&& {\cal R}_{\langle ab\rangle}+
{{1\over3}}\Theta\sigma_{ab}- {{1\over2}}\pi_{ab}-
{{1\over2}}\Pi_{ab}- E_{ab}=0\,, \label{GC}\\ &&{\cal R}-
2(\mud+\mur+\mum)+ {{2\over3}}\Theta^2-2\Lambda=0\,, \label{F}
\end{eqnarray}
where ${\cal R}_{ab}$ and ${\cal R}$ are respectively the
projected Ricci tensor and Ricci scalar. To proceed further we now
assume that the fluid flow remains geodesic (i.e.~$A_a=0$) despite
the magnetic presence. In other words, we impose the force-free
condition on the magnetic field (i.e.~$\varepsilon_{abc}B^b{\rm
curl}B^c=0={{1\over3}}{\rm D}_a\mum+\div\Pi_{a}$).

\section{Calculating the limits}\label{calc}

Our method provides a small refinement of~\cite{MES2}, allowing us
to get slightly stronger limits (by about a factor of about two).
In~\cite{MES2}, limits on $\sigma_{ab}$ were calculated in the
following way: First, Eq.~(\ref{dotpiab}) is solved for
$\sigma_{ab}$, and then the separate limits in Eqs.(7)--(22)
of~\cite{MES2} were inserted to give the following limit:
 \be
\;{|\sigma_{ab}|}{\Theta}<\;83\e_2+\es_2+5\ed_1+\;97\ed_3.
 \ee
However, if, after solving Eq.~(\ref{dotpiab}) for $\sigma_{ab}$,
we use Eq.~(41) of~\cite{MES} to convert
$\pi_{ab}\rightarrow\tau_{ab}$ etc. (i.e.,
$\tau_a\simeq3q_a/4\mur$, $\tau_{ab}\simeq15\pi_{ab}/8\mur$...),
and then use Eqs.~(1)--(4) of~\cite{MES2} (after expanding all
derivatives, and using the relevant evolution equations), some
terms cancel, and we get the tighter limit:
 \be
\;{|\sigma_{ab}|}{\Theta}<\es_2+5\ed_1+\;97\ed_3.\label{sh}
 \ee
This gives simpler limits than using the method in~\cite{MES2}. As
a further example, consider the limits on $E_{ab}$ in the case of
no magnetic field (including cosmological constant):
 \be
\;{|E_{ab}|}{\Theta}<H\brac{10\ed_1
+15\eds_1+2\es_2+3\ess_2+\;{18}{7}\ed_3
 +\;{27}{7}\eds_3} +\;{4}{15}H\Omega_{\mathrm{R}}\e_2,
\label{Elims_first}
 \ee
which is less than the corresponding limit in
Eq.~(28),~\cite{MES2}.

\subsection{Magnetic field strength}

Our first problem is to find limits on $\Pi_{ab}$. In the absence
of magnetic fields, limits on $E_{ab}$ may be found directly from
Eq. ~(\ref{dotsigma}), using Eq.~(\ref{dotpiab}). However, with
magnetic fields present, this will give limits on the combination
$|E_{ab}-\;12\Pi_{ab}|$, so we have to find separate equations for
$E_{ab}$ and $\Pi_{ab}$. We can do this by solving
Eq.~(\ref{dotEab})and the time derivative of Eq.~(\ref{dotsigma}),
which gives
 \ba
\Pi_{ab}&=&-\;{3}{2\Theta}\ddot\sigma_{ab}-\;52\dot\sigma_{ab}
    +\Theta\sigma_{ab}\bra{-\;23+2\;{\mur}{\Theta^2}
    +\;54\;{\mud}{\Theta^2}
    +2\;{\mum}{\Theta^2}-\;{\Lambda}{\Theta^2}}\nonumber\\
    &&~+\;{3}{2\Theta}\dot\pi_{ab}
    +\pi_{ab}
    +\;{3}{4\Theta}\sdel_{\<a}q_{b\>}
    -\;{3}{2\Theta}\curl H_{ab},\label{Pi}
\ea with a similar equation for $E_{ab}$. We have used
Eq.~(\ref{dotPiab}). In the linear regime, we can drop the angled
brackets on time derivatives of PSTF tensors. Decoupling these
quantities has introduced extra uncertainty into our equations, in
the form of $\ddot\sigma_{ab}$. In Eq.~(\ref{Pi}), all the shear
terms may be found in terms of the $\tau_{A_\ell}$ simply by
taking appropriate derivatives of Eq.~(\ref{dotpiab}) and using
Eqs.~(\ref{q->tau}) and (\ref{xi->tau}), followed
by~(\ref{boundsIV}). The only term left to worry about is the
$\curl H_{ab}$ term; however, we may use $\sdel_c$ of
Eq.~(\ref{con3}), given that
 \be
|\sdel_cH_{ab}|<|\sdel_a\sdel_b
\sigma_{cd}|+|\sdel_a\sdel_b\omega_c|.
 \ee
Limits on the second gradient of the shear may be found from
Eq.~(\ref{dotpiab}):
 \be
|\sdel_a\sdel_b\sigma_{cd}| <
\;97H^3\brac{105\eddd_1+14\edd_2+21\edds_2+27\eddd_3}.
 \ee
The rotation term may be found using
$\omega_a=-\curl\sdel_a\mur/2\dot\mur=
\varepsilon_{abc}\sdel^b\sdel^c\mur/8H\mur$ and Eq.~(\ref{dotqa}):
 \be
|\sdel_a\sdel_b\omega_c|<\;{81}{10}
H^3\brac{5\eddd_1+5\eddds_1+6\edddd_2}.
 \ee
So we finally have
 \ba
\frac{|\Pi_{ab}|}{\Theta}&<&H
\left\{\;14\Omega_{\mathrm{M}}\bras{25\ed_1+5\es_2
+\;{45}{7}\ed_3}\right.\nonumber\\
    &+&\left.\Omega_{\mathrm{R}}\bras{9\ed_1
    +\;{8}{15}\e_2+\;{6}{5}\es_2
    +\;{18}{7}\ed_3}
    +\Omega_\Lambda\bras{5\ed_1
    +\es_2+\;97\ed_3}\right.\nonumber\\
    &+&\left.10\ed_1+\;{75}{2}\eds_1
    +\;{45}{2}\edss_1+2\es_2+\;{15}{2}\ess_2
    +\;92\esss_2+\;{18}{7}\ed_3+\;{135}{14}\eds_3
    +\;{81}{14}\edss_3\right.\nonumber\\
    &+&\left.
    \;{117}{4}\eddd_1+\;{27}{4}\eddds_1+3\edd_2+\;92\edds_2+
    \;{81}{10}\edddd_2+\;{81}{14}\eddd_3 \right\};\label{pilims}\\
\frac{|E_{ab}|}
{\Theta}&<&H\left\{\;{1}{8}\Omega_{\mathrm{M}}\bras{25\ed_1+5\es_2
+\;{45}{7}\ed_3}
    +\Omega_{\mathrm{R}}\bras{\;92\ed_1+\;{3}{5}\es_2
    +\;{9}{7}\ed_3}
    \right.\nonumber\\
    &+&\left.\;12\Omega_\Lambda\bras{5\ed_1+\es_2+\;97\ed_3}
    \right.\nonumber\\
    &+&\left.15\ed_1+\;{135}{4}
    \eds_1+\;{45}{4}\edss_1+3\es_2+\;{27}{4}\ess_2
    +\;94\esss_2+\;{27}{7}\ed_3+\;{243}{28}\eds_3
    +\;{81}{28}\edss_3\right.\nonumber\\
    &+&\left. \;{117}{8}\eddd_1+\;{27}{8}
    \eddds_1+\;32\edd_2+\;94\edds_2+
    \;{81}{20}\edddd_2+\;{81}{28}\eddd_3\right\}.
    \label{Elims_last}
 \ea
Comparing Eq.~(\ref{Elims_last}) with the limits found in the
absence of a magnetic field, Eq.~(\ref{Elims_first}), reveals the
extra complexity and uncertainty involved in having just one
additional field.

\subsection{Inhomogeneity of the field}

To find limits on the inhomogeneity, we need to find limits on
$\sdel_a\mum=-3\div\Pi_a$. [Note that solving $\sdel_a$ of
Eqs.~(\ref{Ray}) and (\ref{con4}) for $\sdel_a\mum$ does not work
since it does not separate gradients of $\mum$ and $\mud$.] First
we take $\sdel^a$ of Eq.~(\ref{Pi}), and note that
 \be
\div\curl H_a=\;12\curl\div H_a=-\;14\curl\curl
q_a+\;12\bra{\mud+\;43\mur}\curl\omega_a, \ee so \be |\div\curl
H_a|<\;14|\sdel_a\sdel_b
q|+\;12\bra{\mud+\;43\mur}|\sdel_a\omega_b|.
 \ee
Hence we find that
 \ba
\;{|\sdel_a\mum|}{H^3} &<&{ \max|\sdel_a\mum| \over H^3}
\equiv\Omega_{\mathrm{M}}\brac{\;{2511}{4}\edd_1+\;{243}{8}\edds_1
+54\ed_2+162\eds_2
    +\;{729}{20}\eddd_2
    +\;{2187}{14}\edd_3}\nonumber\\
    &+&\Omega_\Lambda\brac{\;{405}{2}\edd_1+81\ed_2+\;{243}{2}\eds_2
    +\;{729}{14}\edd_3}\nonumber\\
    &+&\Omega_{\mathrm{R}}\brac{972\edd_1
    +\;{81}{2}\edds_1+\;{513}{5}\ed_2+\;{2187}{10}\eds_2
    +\;{243}{5}\eddd_2
    +\;{3645}{14}\edd_3}\nonumber\\
    &+&\;{3645}{2}\edd_1+
    \;{8505}{2}\edds_1+\;{3645}{2}\eddss_1+54\ed_2+567\eds_2
    +972\edss_2 \nonumber\\
    &+&\;{729}{2}\edsss_2+\;{6561}{14}\edd_3+\;{2187}{2}\edds_3
    +\;{6561}{14}\eddss_3.\label{Dmub}
\ea



\vskip 5mm

\begin{thebibliography}{99}

\bibitem{K}
R.M. Kulsrud, in {\em Galactic and Extragalatic Magnetic Fields},
Eds.~R. Beck, P.P. Kronberg and R. Wielebinski (Reidel Dordrecht,
1990); P.P. Kronberg (1994) Rep. Prog. Phys. {\bf 57}, 325; J-L.
Han and R. Wielebinski (2002) Chin. J. Astron. Astrophys. Vol.~2,
No.~4, 293.
\bibitem{GR1}
D. Grasso and H.R. Rubinstein (2001) Phys. Rep. {\bf 348}, 163; L.
Widrow (2002) Rev. Mod. Phys. {\bf 74}, 775;
\bibitem{GR2} D. Grasso and H.R. Rubinstein, Phys. Lett. B (1996)
{\bf 379}, 73; B. Cheng, A.V. Olinto, D.N. Schramm and J.W.
Truran, (1996) Phys. Rev. D {\bf 54}, 4714.
\bibitem{BFS}  J.D. Barrow, P.G. Ferreira, and J. Silk (1997)
Phys. Rev. Lett. {\bf 78}, 3610.
\bibitem{MES3}
R. Maartens, G.F.R. Ellis and W.R. Stoeger (1996) Astron.
Astrophys. {\bf 309}, L7.
\bibitem{CC}
C.A. Clarkson and A.A. Coley (2001) Class. Quantum Grav. {\bf 18},
1305.
\bibitem{SME}
W.R. Stoeger, R. Maartens, and G.F.R. Ellis (1995) Astrophys.~J.
{\bf 443}, 1.
\bibitem{MES}
R. Maartens, G.F.R. Ellis and W.R. Stoeger (1995) Phys. Rev. D
{\bf 51}, 1525.
\bibitem{CL}
A.D. Challinor and A.N. Lasenby (1998) Phys. Rev. D {\bf 58},
023001.
\bibitem{TBM}
C.G. Tsagas and J.D. Barrow (1997) Class. Quantum Grav. {\bf 14},
2539; ibid., (1998) {\bf 15}, 3523; C.G. Tsagas and R. Maartens
(2000) Phys. Rev. D {\bf 61}, 083519.
\bibitem{MES2}
R. Maartens, G.F.R. Ellis and W.R. Stoeger (1995) Phys. Rev. D
{\bf 51}, 5942.
\bibitem{MGE}
R. Maartens, T. Gebbie and G.F.R. Ellis (1999) Phys. Rev. D {\bf
59}, 083506.
\bibitem{SAG}
W.R. Stoeger, M.E. Araujo and T. Gebbie (1999) Astrophys. J. {\bf
476}, 435.




\end{thebibliography}
\end{document}